\documentclass[aip,pop,reprint,amsmath,amssymb]{revtex4-1}

\usepackage{graphicx}
\usepackage{float}
\usepackage{dcolumn}
\usepackage{bm}
\usepackage{color}
\usepackage{epstopdf}

\providecommand{\gtrsim}{\:\raisebox{.25ex}{$>$}\hspace*{-.75em}
\raisebox{-.93ex}{$\sim$}\:}
\providecommand{\lesssim}{\:\raisebox{.25ex}{$<$}\hspace*{-.75em}
\raisebox{-.93ex}{$\sim$}\:}

\newlength{\myfigwidth}
\begin{document}

\title{Power-law scaling of plasma pressure on laser-ablated tin microdroplets}

\author{Dmitry~Kurilovich}
\affiliation{Advanced Research Center for Nanolithography (ARCNL), Science Park 110, 1098 XG Amsterdam, The Netherlands}
\affiliation{Department of Physics and Astronomy, and LaserLaB, Vrije Universiteit Amsterdam, De Boelelaan 1081, 1081 HV Amsterdam, The Netherlands}

\author{Mikhail~M.~Basko}
\affiliation{\mbox{Keldysh Institute of Applied Mathematics, Miusskaya square 4, 125047 Moscow, Russia}}
\affiliation{RnD-ISAN, Promyshlennaya Street 1A, 142191 Troitsk, Moscow, Russia}

\author{Dmitrii~A.~Kim}
\affiliation{\mbox{Keldysh Institute of Applied Mathematics, Miusskaya square 4, 125047 Moscow, Russia}}
\affiliation{RnD-ISAN, Promyshlennaya Street 1A, 142191 Troitsk, Moscow, Russia}

\author{Francesco~Torretti}
\affiliation{Advanced Research Center for Nanolithography (ARCNL), Science Park 110, 1098 XG Amsterdam, The Netherlands}
\affiliation{Department of Physics and Astronomy, and LaserLaB, Vrije Universiteit Amsterdam, De Boelelaan 1081, 1081 HV Amsterdam, The Netherlands}

\author{Ruben~Schupp}
\affiliation{Advanced Research Center for Nanolithography (ARCNL), Science Park 110, 1098 XG Amsterdam, The Netherlands}

\author{Jim~C.~Visschers}
\affiliation{Advanced Research Center for Nanolithography (ARCNL), Science Park 110, 1098 XG Amsterdam, The Netherlands}
\affiliation{Department of Physics and Astronomy, and LaserLaB, Vrije Universiteit Amsterdam, De Boelelaan 1081, 1081 HV Amsterdam, The Netherlands}

\author{Joris~Scheers}
\affiliation{Advanced Research Center for Nanolithography (ARCNL), Science Park 110, 1098 XG Amsterdam, The Netherlands}
\affiliation{Department of Physics and Astronomy, and LaserLaB, Vrije Universiteit Amsterdam, De Boelelaan 1081, 1081 HV Amsterdam, The Netherlands}

\author{\mbox{Ronnie~Hoekstra}}
\affiliation{Advanced Research Center for Nanolithography (ARCNL), Science Park 110, 1098 XG Amsterdam, The Netherlands}
\affiliation{Zernike Institute for Advanced Materials, University of Groningen, Nijenborgh 4, 9747 AG Groningen, The Netherlands}

\author{Wim~Ubachs}
\affiliation{Advanced Research Center for Nanolithography (ARCNL), Science Park 110, 1098 XG Amsterdam, The Netherlands}
\affiliation{Department of Physics and Astronomy, and LaserLaB, Vrije Universiteit Amsterdam, De Boelelaan 1081, 1081 HV Amsterdam, The Netherlands}

\author{Oscar~O.~Versolato}
\email{o.versolato@arcnl.nl}
\affiliation{Advanced Research Center for Nanolithography (ARCNL), Science Park 110, 1098 XG Amsterdam, The Netherlands}
\onecolumngrid
\noindent{Published as D.~Kurilovich~\emph{et~al.}, Phys.~Plasmas \textbf{25}, 012709 (2018). DOI: \href{https://doi.org/10.1063/1.5010899}{10.1063/1.5010899}}

\date{19 January 2018}

\begin{abstract}
The measurement of the propulsion of metallic microdroplets exposed to nanosecond laser pulses provides an elegant method for probing the ablation pressure in dense laser-produced plasma. We present the measurements of the propulsion velocity over three decades in the driving Nd:YAG laser pulse energy, and observe a near-perfect power law dependence. Simulations performed with the RALEF-2D radiation-hydrodynamic code are shown to be in good agreement with the power law above a specific threshold energy. The simulations highlight the importance of radiative losses which significantly modify the power of the pressure scaling. Having found a good agreement between the experiment and the simulations, we investigate the analytic origins of the obtained power law and conclude that none of the available analytic theories is directly applicable for explaining our power exponent.\\
\end{abstract}

\maketitle

\section{Introduction \label{s:intro}}

High-density laser-produced plasmas find many applications, ranging from inertial confinement fusion\cite{Atzeni2004,Craxton2015, Betti2016}, over the propulsion of  small spacecrafts\cite{Phipps2007, Phipps2010}, to sources of extreme ultraviolet (EUV) light for nanolithography\cite{Bakshi2006, Benschop2008, Banine_Koshelev.2011, Fomenkov2017, Mizoguchi2017}. The thermodynamic and radiation transport properties, particularly of high-Z laser-produced plasmas (LPPs), are extremely challenging to measure because of the transient nature of these plasmas, combined with complex equations of state and atomic plasma processes. One thermodynamic variable --- the pressure --- can however be elegantly obtained by measuring the propulsion velocity of metallic liquid microdroplets as a result of a laser-pulse impact\cite{Kurilovich_Klein.2016, Hudgins2016}. In an industrially relevant setting for EUV light production such droplets are irradiated by relatively long ($\sim 10$--100\,ns) laser pulses at modest intensities ($\sim 10^9$--$10^{12}$\,W/cm${}^2$), where the laser absorption takes place mostly through the inverse bremsstrahlung mechanism.

If the pulse length is large compared to the hydrodynamic time scale of the ablation flow, a quasi-stationary regime sets in, where the structure of the ablation front only slowly varies in time. The structure of such quasi-stationary ablation fronts has been extensively studied under various simplifying assumptions for more than 40 years\cite{Kidder1968, Afanasev_Krokhin1972, Afanasev1976, MaCo.82, Mora1982, Atzeni2004, Mulser_Bauer2010, Zhou2015}. However, none of these theoretical works is directly applicable to our system. One of the reasons is the treatment of energy transport by thermal radiation. Another reason is departure from the ideal-gas equation of state (EOS) due to multiple temperature-dependent ionization of the target material. These two effects are of major importance for tin ($Z=50$) targets at the here considered irradiation intensities \cite{Basko_Novikov.2015}. A significant further issue is the non-trivial geometry of the laser-target configuration in our experiments, where a spherical target is irradiated from only one side and an essentially two-dimensional (2D) ablation flow develops. It is likely to alter the scaling laws obtained within one-dimensional (1D) models.

Here, we present measurements of the propulsion velocity of free-falling microdroplets of liquid tin and two of its alloys over three decades in the driving Nd:YAG laser pulse energy, operating at its fundamental wavelength of 1064\,nm. The propulsion velocity is obtained by means of high-resolution stroboscopic shadowgraphy techniques. Our data exhibit a remarkable, near-perfect power law dependence of the propulsion velocity on the laser pulse energy, when allowing for a certain threshold energy below which no propulsion occurs. Further, we provide results of simulations performed with the RALEF-2D \cite{Basko_Maruhn.2009, Basko_Maruhn.2010, Basko_Novikov.2014} radiation-hydrodynamic code and compare these critically to the experimental data. We find a very good agreement between the simulations and the experimental power law in cases well above the threshold energy, but establish a significant disagreement regarding the threshold behavior itself.

Next, we investigate whether the obtained power law can be derived within the conventional approach based on the approximation of a steady-state planar ablation flow but corrected for the strong radiative loss. Interestingly, we conclude that none of the analytic theories available in the literature is directly applicable for explaining the power exponent observed in our experiments. We interpret this as evidence that our scaling  belongs to a more complex class of scalable phenomena. Two- or three- dimensional effects, possibly combined with an essentially non-steady-state behavior, are crucial. Inevitably, the respective power-law exponents can only be calculated by solving numerically an appropriate system of partial differential equations.

\section{Experiment \label{s:EXP}}
\subsection{Experimental setup}

The experimental setup is described in detail in Ref.~\onlinecite{Kurilovich_Klein.2016} and is summarized in the following. Droplets of pure liquid tin (99.995\%), or one of its alloys with indium (50\%) or antimony (5\%), are dispensed from a piezo-driven droplet generator at a repetition rate $\simeq 10$\,kHz with a flight speed of $\simeq 12$\,m/s in a vacuum environment ($\simeq 10^{-7}$\,mbar). The droplets relax to a spherical shape with a fixed initial diameter $D_0$, which slightly varied between different experimental campaigns but stayed in the range $D_0= 2R_0 \approx 45$--47\,$\mu$m, where $R_0$ is the droplet radius. 

The produced droplets pass through the focus of an auxiliary He-Ne laser beam, whose scattered light triggers an injection-seeded Nd:YAG drive laser, operating at a 10-Hz repetition rate. The drive laser pulse, emitted at a $\lambda=$\,1064\,nm wavelength, is circularly polarized and has a Gaussian temporal shape with a duration $t_p = 10.0$\,ns, defined as the full-width at half-maximum (FWHM). By using an appropriate plano-concave lens, the laser beam is focused down to a circular Gaussian spot. The experiments were performed for three different focusing conditions with spot sizes of $d_{foc}= 50$, 100, and 115\,$\mu$m (FWHM). Note that, due to a finite geometrical overlap, the droplets in all cases capture only a fraction of the full laser pulse energy. The pulse energy is varied over three decades, spanning the range 0.15--300\,mJ as measured by using calibrated energy meters, in a manner that does not affect the transversal mode profile of the laser beam.

\begin{figure}[t!]
\includegraphics[scale=1]{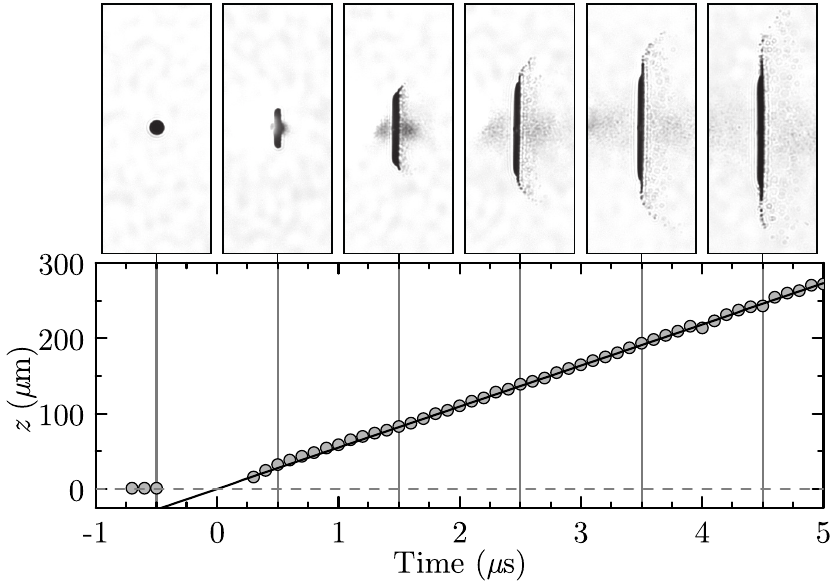}
\caption{\label{f:Fig1} (Top) Stroboscopic side-view shadowgraph images (350\,$\mu$m$\times$800\,$\mu$m) of subsequent tin microdroplets obtained before and after the interaction with a laser pulse. The laser pulse arrives from the left at $t=0$\,$\mu$s. The images represent the case of $E_{od}\approx2$\,mJ, $D_0\approx45$\,$\mu$m and $d_{foc}\approx 100$\,$\mu$m (FWHM). (Bottom) The plot shows the time-dependent position of center-of-pixels of images (circles) along the laser propagation axis $z$ as obtained from the image analysis. The undesired capture of the plasma light causes the disruption of the image analysis at $t\lesssim0.25$\,$\mu$s. Each data point is an average of ten unique images obtained at the same time delay. The solid line shows a linear fit to the data points. The slope of this line corresponds to the propulsion velocity of the microdroplets.}
\end{figure}

The position of the laser-impacted droplet is obtained from shadowgraphs generated by pulsed backlight in combination with long-distance microscopes and CCD cameras. This system provides front-view (at 30$^{\circ}$ with respect to the drive-laser light propagation direction) and side-view (at 90$^{\circ}$) images. By varying the time delay of the backlight pulse with respect to the drive laser pulse, stroboscopic images of consequent droplets are obtained  (see Fig.~\ref{f:Fig1}). The analysis of the images is realized by a code that recognizes the center-of-pixels of the propelled and deformed droplet. Knowing the time delay between the backlight shots with a nanosecond accuracy, the droplet propulsion velocity is obtained from the slope of a linear fit to the time-dependent position of the center-of-pixels.

\subsection{Experimental results}

\begin{figure}[t!]
\includegraphics[scale=1]{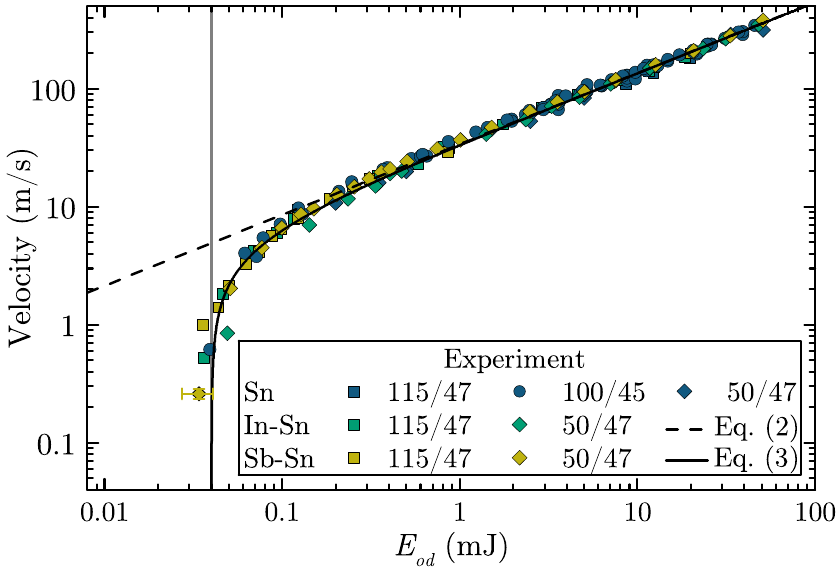}
\caption{\label{f:Fig2_U_EXP_RALEF} Measured propulsion velocity $U$ of Sn, In-Sn, and Sb-Sn droplets as a function of the laser energy $E_{od}$ impinging onto the droplet. The experimental uncertainties have the same values (20--25\% along the $E_{od}$-axis and 10\%  along the $U$-axis) for all measurements. For better visibility the uncertainties are shown only at the lowest laser energy. The focus diameter $d_{foc}$\,($\mu$m) and the droplet diameter $D_0$\,($\mu$m) for different experimental series are indicated in the legend as $d_{foc}/D_0$. The dashed line represents a fit of Eq.~(\ref{Uex:U_pl=}) to the concatenated data for $E_{od}\geq0.2$\,mJ. A fit of Eq.~(\ref{Uex:U_fsc=}) to the full range is depicted as the solid line. The vertical line at $E_{od}= 0.04$\,mJ corresponds to the threshold for droplet propulsion as inferred from this fit.}
\end{figure}

The measured values of the propulsion velocity $U$ are plotted in Fig.~\ref{f:Fig2_U_EXP_RALEF} versus the \emph{energy-on-droplet} $E_{od}$ that is defined as the fraction of the incident laser energy $E$ given by the geometrical overlap of the spatial beam profile in focus and the droplet; in particular, for a Gaussian beam and a spherical droplet we have
\begin{equation}
E_{od} = E\left(1-2^{-D_0^2/d_{foc}^2} \right).
\end{equation}
The thus defined energy-on-droplet appears to be a very convenient parameter, characterizing the effective portion of the laser pulse energy that gives rise to a given value of the propulsion velocity $U$. It also enables the comparison of the results of measurements for different focal spot sizes. As seen from Fig.~\ref{f:Fig2_U_EXP_RALEF}, using this energy parametrization all data fall on a single curve.

Figure~\ref{f:Fig2_U_EXP_RALEF} further demonstrates that, above a certain threshold region of $E_{od,a} \approx 0.1-0.2$\,mJ, the dependence $U(E_{od})$ is well represented by a power law
\begin{equation}\label{Uex:U_pl=}
  U = K_U E_{od}^{\alpha},
\end{equation}
with constant values of the proportionality factor $K_U$\,(m\,s${}^{-1}$mJ${}^{-\alpha}$) and the exponent $\alpha$. A fit of a power law to the full concatenated data set, using the energy range $E_{od} \geq E_{od,a}$, yields $\alpha=$0.60(1). Fitting separately to the individual experimental data sets yields a weighted value of 0.60(1), an identical number, that is bounded by a minimum obtained value of 0.56 and a maximum of 0.63. We note that fitting only the data with a 50-$\mu$m focus size gives a slightly larger power, at 0.62(1). This value, however, is still consistent with the aforementioned result of the fit of the full concatenated data set. Similarly considering only the data from the 100- and 115-$\mu$m size focus cases, yields a power of 0.59(1), consistent with the average of 0.60(1) which is the number used in the comparisons in the following. The value obtained for $K_U$ is, in all cases, consistent with 34(3)\,m\,s${}^{-1}$mJ${}^{-\alpha}$, where the quoted uncertainty is the error in obtaining the absolute magnification of the imaging system.

For $E_{od} < E_{od,a}$, the $U(E_{od})$ curve deviates downward from the simple power law described by Eq.~(\ref{Uex:U_pl=}), with a threshold at $E_{od} =E_{od,0}$.  The parameter range $E_{od,0} < E_{od} < E_{od,a}$ corresponds to a transition regime between the onset of the ablation flow at $E_{od}=E_{od,0}$, and the fully ablative stage at $E_{od} > E_{od,a}$. To incorporate the threshold behavior, the entire set of the experimental points in Fig.~\ref{f:Fig2_U_EXP_RALEF} is fitted by a single shifted power law, defined as
\begin{equation}\label{Uex:U_fsc=}
  U = K_U (E_{od}-E_{od,0})^{\alpha}.
\end{equation}

The value of the offset energy $E_{od,0}$ is obtained by fitting Eq.~(\ref{Uex:U_fsc=}) to the experimental data with $K_U$ and $\alpha$ fixed to the values determined above, i.e. 34\,m\,s${}^{-1}$mJ${}^{-\alpha}$ and 0.60 respectively. The result is shown in Fig.~\ref{f:Fig2_U_EXP_RALEF} and yields a value of $E_{od,0}=0.04(1)$\,mJ. Remarkably, the naive form of Eq.~(\ref{Uex:U_fsc=}) is able to capture all the data to excellent accuracy.

These values are consistent with, and in fact nearly identical to, the values found in our previous work ($\alpha=0.59(3)$, $K_U=35(5)$\,m\,s${}^{-1}$mJ${}^{-\alpha}$, $E_{od,0 }=0.05(1)$\,mJ), dealing with a much smaller data set for solely indium-tin droplets \cite{Kurilovich_Klein.2016}. Consequently, the here demonstrated excellent reproducibility of the data strongly improves the statistical significance of our findings and the broad applicability of the power law. It presents a solid basis for drawing conclusions on the underlying physics.
 
  As is explained in more detail in Section~\ref{s:sim}, the energy $E_{od,a}$ marks the lower boundary of a distinct pattern of laser ablation. In such conditions, the hot plasma with $T\gtrsim 5$--10\,eV envelopes the entire front-illuminated (laser-facing) hemisphere of the droplet, the velocity field across the laser absorption zone approaches that of a quasi-spherical flow, and all the laser flux contributing to $E_{od}$ is efficiently absorbed in the ablated plasma cloud by the inverse bremsstrahlung mechanism. Accordingly, we designate the regime above $E_{od,a}$ as the \emph{fully ablative regime}. In this regime the peak laser intensity on target spans the range $10^9$~W/cm${}^2 < I_l < 3\times 10^{11}$~W/cm${}^2$.
  
\section{Simulation \label{s:sim}}

\subsection{RALEF-2D code}

The simulations reported in this work have been performed with the two-dimensional (2D) radiation-hydrodynamics code RALEF \cite{Basko_Maruhn.2009,Basko_Maruhn.2010}, which has lately been extensively used to simulate laser-driven, droplet-based EUV sources for nanolithography applications \cite{Basko_Novikov.2014, Basko_Novikov.2015, Basko2016}. The hydrodynamics module of RALEF is based on the upgraded version of the \mbox{CAVEAT} package \cite{CAVEAT}, where the second-order Godunov-type algorithm on an adaptive quadrilateral grid is used. The thermal conduction and the spectral radiation transfer (in the quasi-static approximation) are treated within a unified symmetric semi-implicit scheme \cite{LiGl85, Basko_Maruhn.2009} with respect to time discretization. To describe the spatial dependence of the spectral radiation intensity, the classical $S_n$ method is used, combined with the method of short characteristics \cite{DeVo02} to integrate the radiative transfer equation.

The equation of state (EOS) of tin is constructed by using the FEOS model \cite{Faik_Basko.2012} that provides, within a unified model, an adequate and thermodynamically consistent description of high-temperature plasma states together with the low-temperature liquid-gas phase coexistence region. The model for thermal conductivity is based on a semi-empirical expression for the transport cross section of the electron-ion collisions \cite{Basko_Lower.1997}, which enables a smooth matching of the Spitzer plasma conductivity to that of metals near normal conditions.

\begin{figure}[t]
\includegraphics[scale=1]{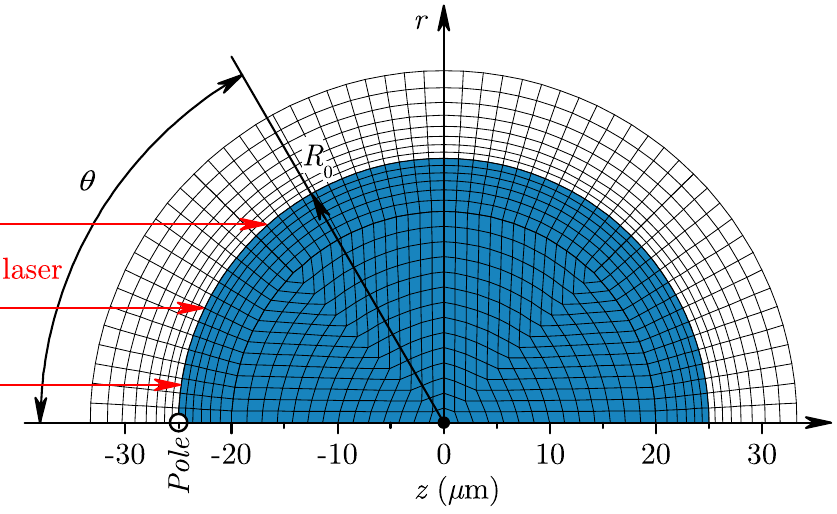}
\caption{\label{f:Fig3} Schematic view of a spherical tin droplet of radius $R_0=25$\,$\mu$m (shaded), projected onto the computational domain with the outer radius of 1\,mm (not shown here) in the $rz$-plane. Depicted is a crude version of the numerical mesh used in the simulation, assuming unpolarized incident laser light.}
\end{figure}

All the simulations are performed for a spherical droplet of pure tin with initial radius $R_0 =25$\,$\mu$m and initial density $\rho_0 =6.9$\,g/cm${}^3$, assuming that slight differences between the physical properties of pure Sn and its two alloys used in the experiments, are insignificant. The adaptive numerical mesh has a topological structure as displayed in Fig.~\ref{f:Fig3}. It extends with 360 zones over the $\pi$ interval of the polar angle $\theta$, and with 350 radial zones over the interval 20\,$\mu$m${} \leq r \leq 1$\,mm. This totals to 142\,200 mesh cells over the simulated half-circle in the $rz$ plane. The mesh is progressively refined in the radial direction towards the droplet surface to resolve the skin layer of the liquid tin. The minimum cell thickness of this layer is 4.5\,nm. The outer region 25\,$\mu$m${} \leq r \leq 1$\,mm is initially filled with a tenuous tin vapor at a density of $\rho_{v0} =10^{-10}$\,g/cm${}^3$.

In all the simulation runs, the same Gaussian temporal power profile of the 1064\,nm laser pulses is used, with the pulse duration $t_p =10$\,ns (FWHM), peaking at $t=1.5 t_p =15$\,ns. The spatial laser profile is also Gaussian, with two values of the focal spot diameter (FWHM): $d_{foc} =115$\,$\mu$m (series A) and $d_{foc} =50$\,$\mu$m (series B). Propagation and absorption of the laser light is calculated within a hybrid model \cite{Basko_Tsygvintsev2017}, which accounts for refraction in the tenuous corona. In addition, it ensures a physically correct description of reflection from the critical surface, including the Fresnel reflection from the metal-vacuum interface. Lastly, the incident light is assumed to be unpolarized.

For all cases in the fully ablative regime, radiative energy transport is important. Radiation generation and transport is treated with the same opacity model as in Ref.~\onlinecite{Basko2016}, where the conversion efficiency into the 13.5-nm EUV emission is investigated for a CO$_2$-laser-driven plasma. The angular dependence of the radiation intensity is modeled with the $S_6$ quadrature, while the spectral dependence is simulated with 28 discrete spectral groups of variable width. Two spectral groups belong to the 2\% band at 13.5\,nm, where the strongest emission from the Sn plasma is expected at sufficiently high laser intensities.

\subsection{Simulation results}

\subsubsection{Droplet propulsion \label{sec:sim_prop}}

\begin{figure}[b]
\includegraphics[scale=1]{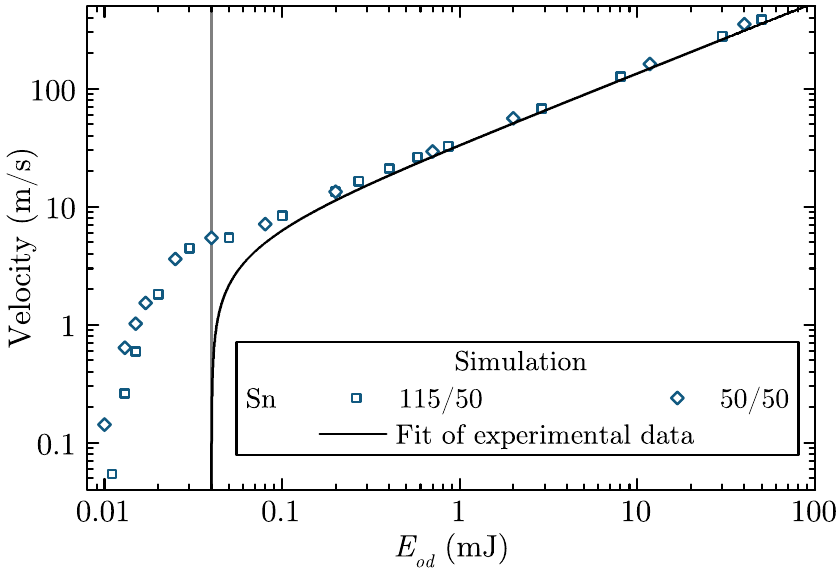}
\caption{\label{f:fig4_U_vs_Eod_RALEF} Dependence of the propulsion velocity $U$ on $E_{od}$ calculated with the RALEF-2D code. The focus diameter $d_{foc}$\,($\mu$m) and the droplet diameter $D_0$\,($\mu$m) for different simulation series are indicated in the legend as $d_{foc}/D_0$.The black curve represents the best fit to the experimental points (see Fig.~\ref{f:Fig2_U_EXP_RALEF}). The vertical line at $E_{od}= 0.04$\,mJ corresponds to the threshold for droplet propulsion as inferred from that fit.}
\end{figure}

The calculated propulsion velocity $U$ for various $E_{od}$ values is plotted in Fig.~\ref{f:fig4_U_vs_Eod_RALEF}. In the RALEF code it is computed as the velocity of the center of mass, comprising all the material with the density in excess of 1\,\% of its maximum value at the time $t=t_f= 200$\,ns. Similarly to the experimental results, for $E_{od} > 0.1-0.2$\,mJ, the dependence $U(E_{od})$ is almost a perfect power law: the deviations of the calculated points from Eq.~(\ref{Uex:U_pl=}) with the best-fit values of
\begin{equation}\label{sim:K_U=}
  K_U =36.0(3)\,\mbox{m\,s${}^{-1}$mJ${}^{-\alpha}$}, \quad \alpha =0.610(5),
\end{equation}

calculated for the combined set of points from series A and B in the range $E_{od} \geq 0.2$\,mJ, do not exceed $\pm 2.5\%$ --- which is practically the intrinsic accuracy of the simulations. Fig.~\ref{f:fig4_U_vs_Eod_RALEF} confirms that within the same $\pm 2.5\%$ accuracy the energy-on-droplet $E_{od}$ proves indeed to be an adequate universal parameter, which unites the $d_{foc}=115$\,$\mu$m and $d_{foc}=50$\,$\mu$m points into virtually a single curve. For the variation of the coefficient $K_U$ with the droplet size $R_0$ and the laser pulse duration $t_p$, we refer to the Appendix.
Judging from Fig.~\ref{f:fig4_U_vs_Eod_RALEF}, the agreement between the calculated and the measured $U$ values in the fully ablative regime could hardly be better: the deviations from the best experimental fit do not exceed 11\%, which lies within the experimental errors. However, the droplet diameter $D_0=50$\,$\mu$m, used in the simulations, slightly exceeds the actual values of $D_0 \approx 45$--47\,$\mu$m. For instance, the correction to a smaller value $D_0=46$\,$\mu$m would raise the calculated $U$ values in the fully ablative regime in Fig.~\ref{f:fig4_U_vs_Eod_RALEF} by some 20\%, leaving the power $\alpha$ unchanged. The fact that the model tends to slightly overestimate the propulsion velocity can, on the one hand, be attributed to a systematic experimental uncertainty, combining possible measurement errors in the spatial beam profile and the droplet diameter. Alternatively, the RALEF simulations may, for example, systematically underestimate the radiation energy losses, whose modeling could still noticeably be improved.

All in all, a very good agreement between the simulation and the experiment is found in the fully ablative regime. Particularly, concerning the scaling exponent~$\alpha$, the best-fit experimental value $\alpha =0.60(1)$ is practically the same as the theoretical value in given Eq.~(\ref{sim:K_U=}). This provides a strong evidence that the RALEF code sufficiently accurately accounts for the key physical processes governing the Sn plasma dynamics in this regime. Therefore, it can be used to extract additional information about the relative role of these processes.

At the low energies $E_{od} < 0.1$\,mJ, the simulation results begin to significantly deviate from the experimental values. Here we have to deal with the initial phase of the onset of ablation, which is controlled by physical processes that are quite distinct from those governing the fully ablative regime. The key role in this initial phase should belong to an adequate modelling of laser-optical properties and propagation of a non-steady thermal wave across a thin surface layer of tin.  In such conditions, this layer is driven into a non-trivial thermodynamic state of superheated metastable liquid, followed by a phase transition into a state of dense hot vapor. We leave the full investigation of this regime for future work.

\subsubsection{Plasma characterization in the fully ablative regime}

\begin{figure}[t]
\includegraphics[scale=1]{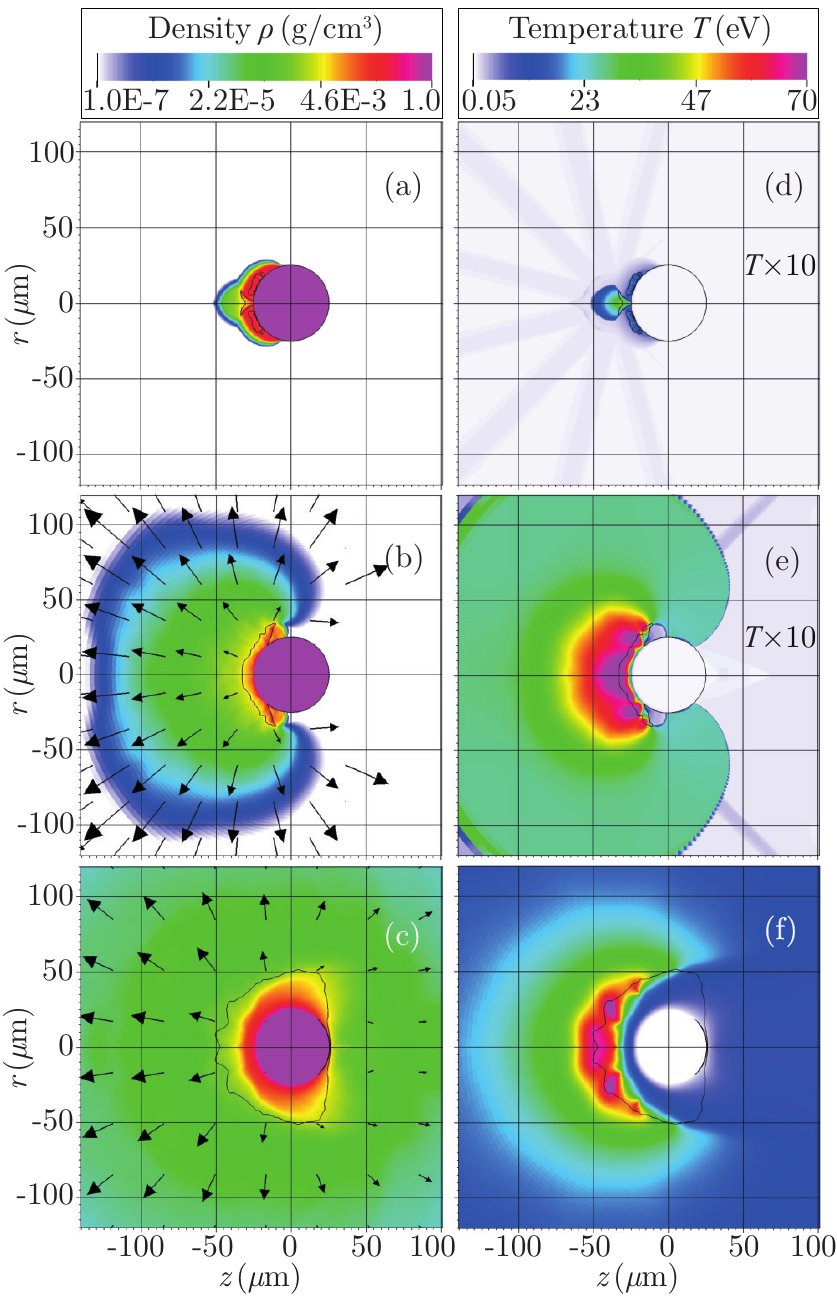}
\caption{\label{f:fig5_2D}  Calculated 2D density and temperature color maps for the cases $E_{od} = 0.06$\,mJ (a, d), $0.2$\,mJ (b, e), and $30$\,mJ (c, f) $d_{foc}= 115$\,$\mu$m  at $t=15$\,ns when the laser power peaks. The black curve is the isocontour of the free electron density $n_e =0.1n_{cr} = 10^{20}$\,cm${}^{-3}$. Black arrows in (b, c) indicate the velocity field in the outflowing plasma.}
\end{figure}

A general perception of the plasma dynamics in the fully ablative regime can be obtained from Fig.~\ref{f:fig5_2D}, which displays the 2D density and temperature distributions for the two cases of $E_{od}= 0.2$\,mJ and 30\,mJ at time $t=15$\,ns, coinciding with peak laser power.
As is seen in Figs.~\ref{f:fig5_2D}(b) and (c), a characteristic feature of the fully ablative regime is a stabilized geometry of the plasma flow across the laser absorption zone. The latter manifests itself in Figs.~\ref{f:fig5_2D}(e) and (f) as the region with highest plasma temperatures. Note that the peak temperature in the ablative regime varies with $E_{od}$ over a wide range of 5\,eV${} \lesssim T \lesssim 100$\,eV.  In all cases with $E_{od}\geq E_{od,a}$, by the middle of the pulse, the plasma plume attains a size of several $R_0$ and occupies the entire $2\pi$ of the solid angle above the illuminated droplet hemisphere; the velocity field stabilizes to a quasi-steady, quasi-spherically diverging pattern; the laser-absorption zone itself reaches its maximum size, which becomes practically independent of $E_{od}$.

Intuitively it is clear that, once the 2D (or 3D) geometry of the plasma flow and laser absorption settles down to a stable pattern, the principal ablation parameters (like the characteristic pressure, temperature, ablation velocity, etc.) can be expected to become scalable. On the other hand, in the low-energy cases with $E_{od} < E_{od,a}$ (see Figs.~\ref{f:fig5_2D}(a) and (d)), intense laser absorption takes place in a narrow plasma plume near the target pole while a large portion of the incident flux contributing to $E_{od}$ is reflected from a cooler and sharper liquid-vapor boundary at $\theta \gtrsim 40^{\circ}$--$50^{\circ}$. Therefore, the ablation parameters from these low-energy cases cannot be expected to be scalable in the same way as those in the fully ablative regime.

\begin{figure}[b]
\includegraphics[width=\myfigwidth]{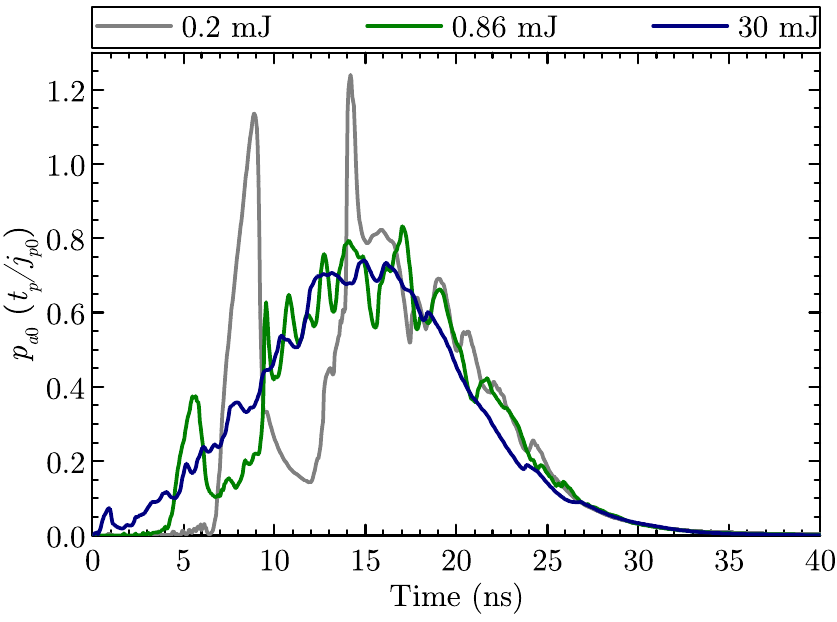}
\caption{\label{f:fig6_pa0_vs_t} Calculated temporal dependence of the ablation pressure at the droplet pole $p_{a0}(t)$ normalized by the quotient $t_p/j_{p0}$ of the laser pulse length and the pressure impulse  for three values of $E_{od}$ and  $d_{foc}=115$\,$\mu$m.}
\end{figure}

Fig.~\ref{f:fig5_2D} also demonstrates that the ablation flow is subject to hydrodynamic instabilities, the most salient of which appears to be the self-focusing instability due to laser refraction in the underdense plasma, inherent in the laser deposition model \cite{Basko_Tsygvintsev2017}. The resulting irregular fluctuations of the plasma parameters in space and time manifest themselves as ``spotty''  temperature distributions and ``wavy''  $n_e$ isocontours in Fig.~\ref{f:fig5_2D}. The temporal variation of the ablation pressure at a fixed location, illustrated in Fig.~\ref{f:fig6_pa0_vs_t} for the target pole, becomes especially violent for low $E_{od}$ values.

Although the self-focusing instability has a clear physical origin, the amplitude of the ensuing fluctuations tends to be overestimated in the present RALEF simulations (especially on length scales comparable to, or smaller than the laser wavelength $\lambda$) due to the absence of diffraction effects in the laser propagation model \cite{Basko_Tsygvintsev2017}. However, when averaged over space and time, the impact of this ``noise'' on the calculated $U$ values turns out to be negligible, i.e.\ on the level of $\pm 1\%$ as ascertained by dedicated computer runs. Having verified it in 2D, we expect no more than only a moderate, by about a factor of 1.5, increase of this effect in the full 3D approach. This is similar to what has firmly been established for the nonlinear stage of the Rayleigh-Taylor instability \cite{Kull_1991}.

\subsubsection{Ablation pressure \label{s:aP}}

The ablation-plasma parameter most directly related to the propulsion velocity $U$ is the ablation pressure. More specifically, the velocity $U$ can be determined from the relationship
\begin{equation}\label{aP:MU=}
  MU=P,
\end{equation}
where $M$ is the total mass, and $P$ is the total momentum of the liquid tin at a certain moment $t_f \gg t_p$. As the entire simulated configuration is axisymmetric, the total momentum vector $P$ lies along the $z$ axis. In our case, the results become insensitive to $t_f$ for $t_f \gtrsim 100$~ns, thus we present results for $t_f=200$\,ns. From the simulations we learn that the ablated mass fraction $\delta_M$, defined as the relative fraction of the total tin mass with $\rho <0.1$\,g/cm${}^3$, does not exceed 10\% for the entire range of $E_{od} \leq 40$~mJ (see Tables~\ref{t:tab-A} and \ref{t:tab-B}). The subsequent deformation of the ablated surface is not significant (see Fig.~\ref{f:fig5_2D}). Then, the propulsion momentum $P$ can be evaluated as
\begin{equation}\label{aP:P=}
  P= 2\pi R_0^2 \int\limits_{0}^{\pi} j_p(\theta) \sin\theta \cos\theta\, d\theta, \quad j_p(\theta) = \int\limits_{0}^{t_f} p_a(t,\theta)\, dt,
\end{equation}
where $p_a(t,\theta)$ is the ablation pressure at the spherical droplet surface as a function of time $t$ and polar angle $\theta$, and $j_p(\theta)$  is the local impulse of the ablation pressure. Note that $\theta$ is measured with respect to the negative direction of the rotation axis $z$, as is shown in Fig.~\ref{f:Fig3}.
\begin{table} [b]
\caption{\label{t:tab-A} Calculated ablation parameters (propulsion velocity $U$, ablated mass fraction $\delta_M$, radiative loss fraction $\phi_r$, laser absorption fraction $f_{la,od}$, spatial form-factor of ablation pressure $\langle\bar{j}_{p\theta}\rangle$) for a selection of laser energies with $d_{foc} =115$\,$\mu$m.}

\begin{ruledtabular}
\begin{tabular}{lccccc}
 $E_{od}$ (mJ) &  0.2 & 0.86 & 2.88 &  8.06 & 30 \\ 
 $U$ (m/s) & 13.5 & 32.7 & 67.4 & 128 & 280 \\ \hline
 $\delta_M$ & 0.006 & 0.009 & 0.016 & 0.034 & 0.085  \\
 $\phi_r$ & 0.23 & 0.35 & 0.51 & 0.64 & 0.74  \\
 $f_{la,od}$ & 0.83 & 0.91 & 0.93 & 0.92 & 0.96 \\
 $\langle\bar{j}_{p\theta}\rangle$ & 0.595 & 0.571 & 0.567 & 0.580 & 0.585 \\
\end{tabular}
\end{ruledtabular}

\bigskip

\caption{\label{t:tab-B} Same as Table~\ref{t:tab-A} but for $d_{foc} =50$\,$\mu$m.\hspace{1.6cm}}
\begin{ruledtabular}
\begin{tabular}{lccccc}
 $E_{od}$ (mJ) &  0.2 & 0.7 & 2.0 &  11.75 & 40 \\ 
 $U$ (m/s) & 13.4 & 29.4 & 56.2 & 162 & 354 \\ \hline
 $\delta_M$ & 0.006 & 0.008 & 0.013 & 0.042 & 0.093  \\
 $\phi_r$ & 0.22 & 0.33 & 0.44 & 0.63 & 0.69  \\
 $f_{la,od}$ & 0.77 & 0.89 & 0.94 & 0.93 & 0.97 \\
 $\langle\bar{j}_{p\theta}\rangle$ & 0.503 & 0.508 & 0.508 & 0.529 & 0.568 \\
\end{tabular}
\end{ruledtabular}
\end{table}

Equations (\ref{aP:MU=}) and (\ref{aP:P=}) can be used to relate the established scaling of $U$ with $E_{od}$ in Fig.~\ref{f:Fig2_U_EXP_RALEF} to existing analytic scaling laws for the ablation pressure $p_a$. However, all the previous analytic results on the scaling of $p_a$ with the incident laser flux $I_l$ have been obtained under a few assumptions.  It is assumed that the ablation flow either (i)~has a 1D planar geometry ($p_a$ is constant in space), or (ii)~is in a steady state ($p_a$ is independent of time), or both \cite{Mulser_Bauer2010}. Unfortunately, neither of these assumptions can be considered as adequate for our situation. Nonetheless, the effects of the spatial, along the droplet surface, and the temporal variations of the ablation pressure $p_a(t,\theta)$ can be separated as follows.

One can rewrite Eq.~(\ref{aP:P=}) as
\begin{equation}\label{aP:P==}
  P =\pi R_0^2 \,j_{p0}\, \langle\bar{j}_{p\theta}\rangle, \quad j_{p0} \equiv j_p(0)=\int\limits_{0}^{t_f} p_a(t,0)\, dt,
\end{equation}
where
\begin{equation}\label{aP:<jtet>=}
  \langle\bar{j}_{p\theta}\rangle = 2\int\limits_{0}^{\pi} \bar{j}_p(\theta) \sin\theta \cos\theta\, d\theta, \quad \bar{j}_p(\theta) \equiv j_p(\theta)/j_{p0}.
\end{equation}
Our simulations demonstrate that in the fully ablative regime the dimensionless \emph{spatial form-factor} $\langle\bar{j}_{p\theta}\rangle$ of the pressure impulse barely depends on the incident laser flux when the focal spot is fixed (see Tables~\ref{t:tab-A} and \ref{t:tab-B}). For $d_{foc} =115$\,$\mu$m, for instance, it fluctuates in the range $\langle\bar{j}_{p\theta}\rangle \approx$ 0.57--0.59, remaining virtually constant within our simulation accuracy.
Hence, as long as we can neglect small variations of the mass $M$ and size $R_0$ of the irradiated droplet, the problem of the analytic derivation of the scaling of $U$ with $E_{od}$ is reduced to the derivation of the analogous scaling for the local (at the pole) pressure impulse $j_{p0}$. Before tackling this issue, we provide some additional information on the angular dependence of the ablation pressure that might be helpful for a general analysis of the hydrodynamic response of liquid droplets to laser pulses \cite{Gelderblom_Lhuissier.2016, Kurilovich_Klein.2016}.

\begin{figure}[t]
\includegraphics[width=\myfigwidth]{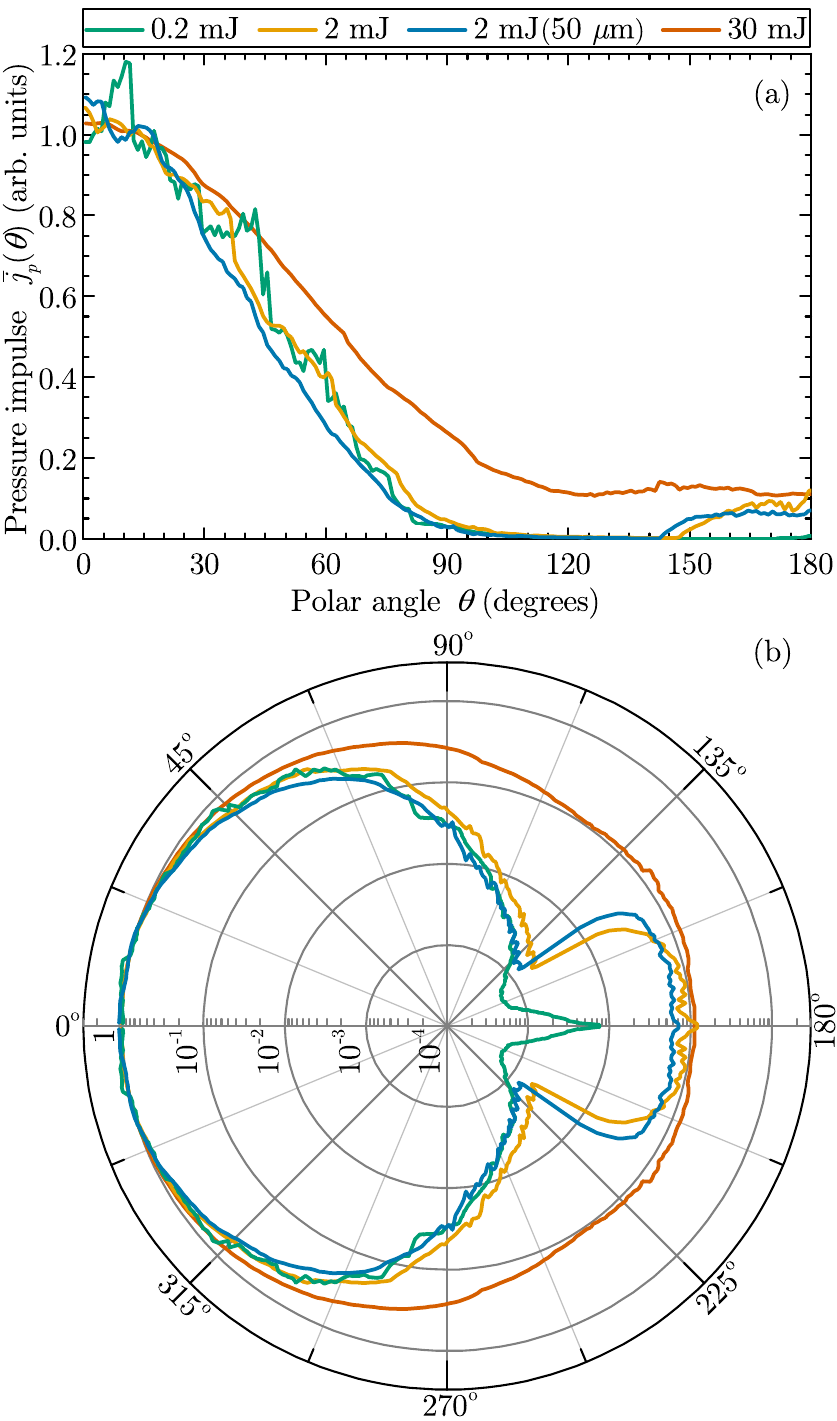}
\caption{\label{f:fig7_jp_vs_tet} (a) Calculated variation of the normalized pressure impulse $\bar{j}_p(\theta)$ along the surface of the spherical droplet. The polar angle $\theta$ is measured relative to the direction towards the drive laser. Shown are three cases with $E_{od} = 0.2$, 2.0, and 30\,mJ for the focal spot $d_{foc}=115$~$\mu$m, and, for comparison, one case with $E_{od} = 2.0$\,mJ for $d_{foc}=50$\,$\mu$m. (b) Same as (a) but in the polar plot representation with the radial coordinate in logarithmic scale.}
\end{figure}

Fig.~\ref{f:fig7_jp_vs_tet} shows several angular profiles of the normalized pressure impulse $\bar{j}_p(\theta)$, calculated with the RALEF code. Despite the fact that the $\bar{j}_p(\theta)$ curve for the highest-energy case $E_{od}=30$\,mJ is clearly broader than those for lower pulse energies, its integral (see~Eq.~(\ref{aP:<jtet>=})) remains practically the same because of the negative contribution from the backward hemisphere $\theta >90^{\circ}$. A salient local rise of $\bar{j}_p(\theta)$ at $\theta \gtrsim 150^{\circ}$ for the 2-mJ case is explained by the plasma flowing around the droplet and accumulating on its horizontal axis. It leaves a local cloud of relatively dense and hot vapor, which exerts a noticeable backward pressure onto the droplet for some 30--50\,ns after the laser has already been off. We further note that, for the same $E_{od}=2$\,mJ, a tighter laser focus (the $d_{foc} =50$\,$\mu$m curves) produces an only slightly narrower pressure profile $\bar{j}_p(\theta)$.

\section{Analytic scaling laws \label{s:as}}

Having found an excellent agreement between the experiment and simulations, we will attempt to derive the obtained scaling law analytically on the basis of an appropriately simplified model. Additional information, available from the simulations, provides guidance for working out such a model.

Analytic scaling laws are usually derived for the ablation pressure $p_a$ as a function of the \emph{hydrodynamically absorbed flux} $I_{lh}$ (W/cm${}^2$), assumed to be constant in time and fully converted into the kinetic and internal energy of the ablated material \cite{Mulser_Bauer2010}. To simplify the argumentation, we focus our attention on the simulations (series~A) with a fixed spot size $d_{foc}=115$\,$\mu$m. Then, because all the pulses have the same temporal profile, the polar incident flux $I_{l,0}(t)$, the incident laser energy $E$, and the energy-on-droplet $E_{od}$ are all directly proportional to one another, as well as to the polar energy fluence $F_{l,0}=\int I_{l,0}(t)\, dt$. Consequently, an approximate analytic scaling of $U$ with $E_{od}$ could be obtained by (i)~relating the incident laser fluence $F_{l,0}$ to the hydrodynamically absorbed one $F_{lh,0}$, and (ii)~making an assumption that the time-integrated quantities $j_{p0}$ and $F_{lh,0}= \int I_{lh,0}(t)\, dt$ scale with one another in the same way as $p_a$ and $I_{lh}$ in a steady-state planar 1D ablation front, for which analytic results are available. Here we assume that the droplet mass $M$ and the 2D form-factor $\langle\bar{j}_{p\theta}\rangle$ in Eqs.~(\ref{aP:MU=}) and (\ref{aP:P==}) are constant. Note that assumption (ii) is by no means obvious, and might, in fact, be rather inaccurate.

\subsection{Laser absorption and radiative losses}

There are two main loss mechanisms that reduce the incident laser energy fluence $F_{l,0}$ to the hydrodynamically absorbed one $F_{lh,0}$, namely, partial reflection of the laser light and radiative losses. Accordingly, since $F_{l,0}$ is directly proportional to $E_{od}$, we can, following our logic, introduce a \emph{hydrodynamically absorbed energy-on-droplet}
\begin{equation}\label{as:E_odh=}
  E_{od,h} = f_{la}(1-\phi_r)E_{od}.
\end{equation}

In Eq.~(\ref{as:E_odh=}) $f_{la}$ is the laser energy absorption fraction, and $\phi_r$ is the fraction of the absorbed laser energy which escapes from the plasma by thermal emission.
Having introduced effective corrections for the laser reflection and radiative losses by means of Eq.~(\ref{as:E_odh=}), we take the next step and relate the resulting scaling of $j_{p0}$ with $E_{od,h}$ to an analytic scaling of $p_a$ with $I_{lh}$ predicted by an appropriate 1D model. If a close agreement were found, we could accept the invoked 1D model as an appropriate one for the interpretation of our experiments.

Strictly speaking, both factors $f_{la}$ and $(1-\phi_r)$ in Eq.~(\ref{as:E_odh=}) must be calculated at the target pole. But even a simplest analytic model for evaluating $f_{la}$ and $\phi_r$ would be too cumbersome for the present work \cite{Basko_Novikov.2015}. Instead, we take their values from the RALEF simulations. The problem, however, is that the local polar value of $\phi_r$ cannot be extracted from the simulations. Moreover, it is an ill-defined quantity because of the non-local nature of radiation transport. Thus, we are forced to use the integral values of $\phi_r$, calculated for the whole plasma volume and listed in Tables~\ref{t:tab-A} and \ref{t:tab-B}. For the laser absorption, whose impact on the scaling is considerably less important ($\Delta\alpha \approx 0.03$), we also use the integral values of $f_{la} = f_{la,od}$, calculated for the laser energy fluence over the cross-section $\pi R_0^2$ of the droplet. These values are consistent with the integral values of $\phi_r$ and exhibit weaker instability variations than the local polar values $f_{la,0}$.

First of all we note that the calculated values of $\phi_r$, ranging from $ \simeq 20\%$ to $\gtrsim 70\%$ as $E_{od}$ increases from 0.2\,mJ to 40\,mJ, provide clear evidence of the important role played by radiative losses in our situation. For the scaling exponent it is important that the coefficient $(1-\phi_r)$ changes by about a factor of 2.5--3 over the considered range of $E_{od}$, which implies an exponent shift by $\Delta\alpha \approx 0.17$.

\begin{figure}[t]
\includegraphics[width=\myfigwidth]{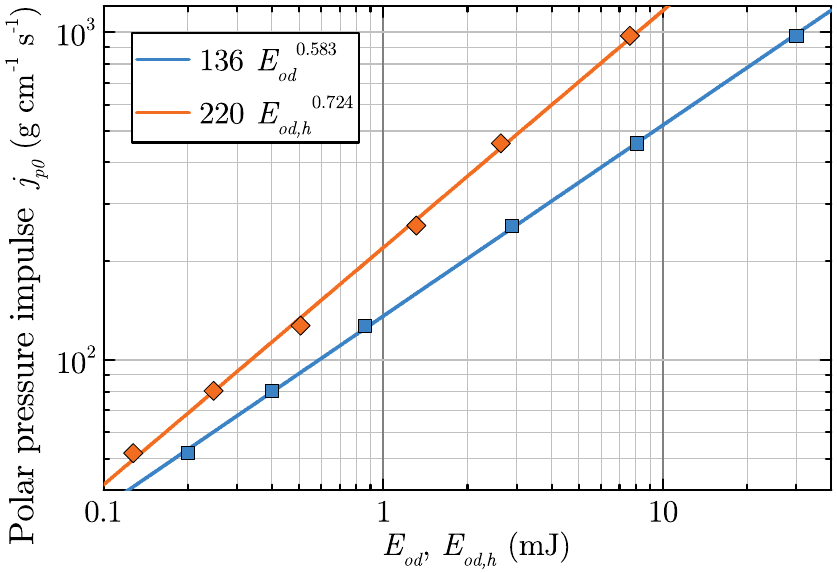}
\caption{\label{f:fig8_jp0_vs_E} Calculated pressure impulse $j_{p0}$ at the illuminated droplet pole as a function of the energy-on-droplet $E_{od}$ for the $d_{foc}=115$\,$\mu$m case, and of the radiatively-corrected energy-on-droplet $E_{od,h}$. }
\end{figure}

Fig.~\ref{f:fig8_jp0_vs_E} shows the dependence of the calculated pressure impulse $j_{p0}$ on the incident, $E_{od}$, and hydrodynamically absorbed, $E_{od,h}$, energy-on-droplet. Solid lines represent the respective power-law fits, that yield the following exponents.
\begin{equation}\label{as:jp0-prop=}
  j_{p0} \propto E_{od}^{0.583 \pm 0.005} \propto E_{od,h}^{0.724 \pm 0.014}.
\end{equation}
The results of the fits significantly differ from one another. This difference of $\Delta\alpha \approx 0.14$ provides a quantitative measure of the influence of radiative losses on the discussed scaling law. In fact, this influence is even stronger ($\Delta\alpha \approx 0.17$) since the two factors $f_{la}$ and $(1-\phi_r)$ in Eq.~(\ref{as:E_odh=}) change in opposite directions (see Tables~\ref{t:tab-A} and \ref{t:tab-B}).
Clearly, it is the second exponent $\alpha = 0.724(14)$ that should be compared with the known analytic scalings for $p_a(I_{lh})$. A noticeably larger statistical uncertainty in this exponent ($\pm 0.014$ versus $\pm 0.005$, thus comparable to the experimental error), related to the goodness of fit, is apparently caused by using the integral values of $\phi_r$ and $f_{la}$, which ``feel'' the 2D ablation geometry of a spherical droplet.

Note that the exponent $\alpha =0.583(5)$ for the $j_{p0}(E_{od})$ dependence differs slightly from the previously quoted value of $\alpha =0.610(5)$  for the $U(E_{od})$ scaling (see Section~\ref{sec:sim_prop}). This difference of $\Delta\alpha \approx 0.03$ arises from the fact that the remaining liquid mass $M$ in Eq.~(\ref{aP:MU=}) decreases by about 9\% as $E_{od}$ increases from 0.2\,mJ to 30\,mJ, and less impulse is needed to attain a given velocity $U$.
\subsection{Effects of the equation of state}

Well-known theoretical models of 1D quasi-stationary ablation fronts, based on the ideal-gas equation of state (EOS) with the adiabatic index $\gamma=5/3$, yield two limiting scaling laws for the ablation pressure. Namely, the one for the case where laser absorption occurs in an infinitely thin layer at the critical surface \cite{Kidder1968,MaCo.82,Mora1982} (case~I), and the other one for the case where laser light is absorbed in an extended region by the inverse bremsstrahlung mechanism before reaching the critical surface \cite{Nemchinov1967, Basov_Gribkov.1968, Mora1982} (case~II),
\renewcommand{\arraystretch}{1.5}
\begin{equation}\label{as:idealgas-pa-prop(I-II)=}
  p_a \propto \left\{ \begin{array}{ll} I_{lh}^{2/3}, & \mbox{case I (ideal-gas EOS)}, \\
  I_{lh}^{7/9} L^{-1/9}, & \mbox{case II (ideal-gas EOS)}.
  \end{array} \right.
\end{equation}
\renewcommand{\arraystretch}{1}
In case~II, an additional relevant parameter enters the scaling, which is the density-gradient length $L$ in the absorption zone. For quasi-spherical (or cylindrical) diverging flows, where a steady-state solution with a sonic point exists, $L$ should be set equal to the radius of the sonic point \cite{Nemchinov1967}. In the planar geometry, where no steady-state solution is possible \cite{Nemchinov1967}, one can assume the laser to be absorbed in a non-steady rarefaction wave in expanding plasma, where $L \propto c_st$, and $c_s$ is the characteristic sound velocity. In this way one arrives at yet another well-known analytic scaling $p_a \propto I_{lh}^{3/4} t^{-1/8}$, applicable to non-steady planar ablation flows with the ideal-gas EOS \cite{Nemchinov1967,Kidder1968,Caruso_Gratton1968,Mora1982}.

All the above analytic scalings with rational-number exponents, based on the ideal-gas EOS, can definitely be applied to interpretation of experiments on low-Z targets (like plastic foils) that are fully ionized by a sufficiently high laser energy flux. None of them, however, can be employed in our case, where a temperature-dependent ionization of tin ($Z=50$) changes the appropriate planar analytic scalings in Eq.~(\ref{as:idealgas-pa-prop(I-II)=}) to \cite{Basko_Novikov.2015}

\renewcommand{\arraystretch}{1.5}
\begin{equation}\label{as:pa-prop(I-II)=}
  p_a \propto \left\{ \begin{array}{ll} I_{lh}^{0.56}, & \mbox{case I (Sn EOS)}, \\
  I_{lh}^{0.64} L^{-0.18}, & \mbox{case II (Sn EOS)}.
  \end{array} \right.
\end{equation}
\renewcommand{\arraystretch}{1}
The experimental situation analyzed here lies between these two cases but closer to case~II. We compare the exponent $\alpha =0.724(14)$ in Eq.~(\ref{as:jp0-prop=}) with $0.56 \lesssim \alpha \lesssim 0.64$ in Eq.~(\ref{as:pa-prop(I-II)=}). The effect of variation of the density-gradient scale $L$ with the laser intensity $I_{lh}$ for case~II is small and only enhances the discrepancy because $L$ can only grow with $I_{lh}$. From comparison between Figs.~\ref{f:fig5_2D}(b) and (c) one infers that the radius of the absorption zone increases by no more than a factor of 1.7 as $E_{od}$ increases from 0.2\,mJ to 30\,mJ, implying  an effective reduction of the scaling exponent by $\Delta\alpha \approx -0.02$.

Thus, a good agreement with the appropriate analytical scaling could have been claimed if Fig.~\ref{f:fig8_jp0_vs_E} demonstrated $j_{p0} \propto E_{od,h}^{\alpha}$ with $0.56 \lesssim \alpha \lesssim 0.62$ --- which is obviously not the case. A superficial observation that the scaling~(\ref{as:jp0-prop=}) of $j_{p0}$ with $E_{od,h}$ is very close to the theoretical result $p_a \propto I_{lh}^{3/4}$ (with $t \approx t_p$ being fixed) should be qualified as incidental. Summarizing, we conclude that the scaling~(\ref{Uex:U_pl=}), (\ref{sim:K_U=}) of the propulsion velocity $U$ with the energy-on-droplet $E_{od}$, established in this work, cannot be derived from the previously published 1D analytic models of the laser ablation fronts.

\section{Conclusion}

Having performed an extensive series of experiments with Nd:YAG laser pulses at different focusing conditions, we have found that within a certain range of laser-pulse energies, covering more than three decades in magnitude, the propulsion velocity of tin droplets scales as a power law $U \propto E_{od}^{\alpha}$ of the energy-on-droplet $E_{od}$ (the incident laser energy intercepted by the cross-section of the droplet). The theoretical analysis, based on 2D simulations with the radiation-hydrodynamics RALEF code, has revealed that the scalability range corresponds to a fully developed regime of laser ablation, where the zone of laser absorption (by inverse bremsstrahlung) in the ablated plasma settles to a stable configuration. For droplets with radii $R_0 \approx 25$\,$\mu$m it starts at $E_{od} \gtrsim 0.1$--0.2\,mJ. The scaling exponent $\alpha =0.610(5)$, obtained from the RALEF results, agrees perfectly with the experimental value of $\alpha =0.60(1)$. The performed analysis demonstrates how the propulsion of metallic microdroplets by a laser-pulse impact can be a good probe for the plasma ablation pressure.

It should be noted that our study was done under a rather unique combination of conditions. A spherical target composed of a high-Z material was irradiated from one side and propelled by an essentially 2D ablation flow. Since the vast majority of previous measurements of the laser ablation pressure were done on low-Z planar targets or on pellets with spherically symmetric irradiation geometry (see, e.g., [\onlinecite{Goldsack1982,Maaswinkel1984,Eidmann1984,Dahmani1992,Dahmani1993}]), we chose to avoid direct comparison of our results to those obtained in these other works, as spurious coincidence of two numbers from different experiments could obfuscate the underlying physics. Instead, we focused our efforts on analyzing the main physical effects that determine our scaling power.

A thorough examination, facilitated by additional information from the RALEF simulations, of the physical processes governing the fully ablative regime in our series of experiments has revealed that the scaling law cannot be directly derived from any of the existing analytic models of quasi-steady 1D ablation fronts. Moreover, this cannot be done even after the effects of radiation energy losses and realistic EOS of tin have been accounted for. The cause must be a complex, essentially 2D (or even 3D) structure of the ablation plasma flow, where the non-local energy transport by thermal radiation in both lateral and radial directions plays an important role. An additional complication comes from the finite pulse length $t_p= 10$\,ns. It is difficult to justify the steady-state approximation, usually implied by analytic evaluation of the scaling exponent, when $t_p$ remains fixed. While the timescale of flow relaxation \cite{Basko_Novikov.2015} to a quasi-steady state is comparable with $t_p$ at $E_{od} = 0.2$\,mJ, it decreases by about a factor of 3--4 at the upper end $E_{od} =30$--50\,mJ of the explored range.

In conclusion, the established scaling of the plasma-propulsion velocity $U$ of tin microdroplets with laser energy $E_{od}$ belongs to a class of scaling laws where theoretical evaluation of the scaling exponent requires the numerical solution of partial differential equations that capture the relevant physical effects in two or three dimensions.

\begin{acknowledgments}
Part of this work has been carried out at the Advanced Research Center for Nanolithography (ARCNL), a public-private partnership between the University of Amsterdam (UvA), the Vrije Universiteit Amsterdam (VU), the Netherlands Organization for Scientific Research (NWO) and the semiconductor equipment manufacturer ASML. We acknowledge the assistance of the mechanical workshop and the design, electronics, and software departments of AMOLF as well as the direct technical support at ARCNL. We also thank SURFsara (\mbox{\url{www.surfsara.nl}}) for the support in using the Lisa Compute Cluster. This work was partially (the contributions by M.~M.~Basko and D.~A.~Kim) funded by the Russian Science Foundation through grant No.~14-11-00699-$\Pi$.
\end{acknowledgments}

\appendix*

\section{Dependence of the propulsion velocity on the droplet size and laser pulse duration}

Having established the scaling Eqs.(\ref{Uex:U_pl=}) and (\ref{sim:K_U=}) of the propulsion velocity $U$ with the energy-on-droplet $E_{od}$, one can, following the logic of Section~\ref{s:aP} and making some reasonable assumptions, evaluate the dependence of $U$ on the droplet radius $R_0$ and the laser pulse duration $t_p$. This might be useful for practical applications.

First of all, we suppose that the exponent $\alpha$ in Eq.~(\ref{Uex:U_pl=}) does not vary with $R_0$ and $t_p$, and only the dimensional coefficient $K_U$ changes. If, when varying $R_0$, we keep the values of the polar energy fluence $F_{l,0}=\int I_{l,0}(t)\, dt$ and of the ratio $R_0/d_{foc}$ fixed, both the polar pressure impulse $j_{p0}$ and the form-factor $\langle\bar{j}_{p\theta}\rangle$ should remain practically unchanged. Then, having noted that in Eq.~(\ref{aP:MU=}) $M \propto R_0^3$ and, as it follows from Eq.~(\ref{aP:P==}), $P \propto R_0^2$, we obtain $U = K_UE_{od}^{\alpha} \propto R_0^{-1}$. Finally, because for fixed $F_{l,0}$ and $R_0/d_{foc}$ one has $E_{od} \propto R_0^2$, we arrive at
\begin{equation}\label{app:K_U-R0=}
  K_U \propto R_0^{-1-2\alpha}.
\end{equation}

Similarly, we can deduce the scaling with the pulse duration $t_p$ by assuming that the Gaussian pulse profile is simply stretched in time by a factor $a$ ($t_p \rightarrow at_p$), with the peak laser intensity kept fixed. Then, because the local (polar) ablation pressure $p_a(t,0)$ depends primarily on the local laser intensity, one can surmise that the corresponding pressure pulse will also be simply stretched in time by the same factor $a$. As a result, the propulsion velocity would scale as $U\rightarrow aU$. Since $E_{od}$ in Eq.~(\ref{Uex:U_pl=}) is directly proportional to $t_p$, the factor $K_U$ should scale as

\begin{equation}\label{app:K_U-tp=}
  K_U \propto  t_p^{1-\alpha}.
\end{equation}

Finally, rounding off the $K_U$ and $\alpha$ values from Eq.~(\ref{sim:K_U=}), we obtain
\begin{equation}\label{app:K_U=}
  K_U  \approx 36 \left(\frac{\mbox{25~$\mu$m}}{R_0} \right)^{2.2} \left(\frac{t_p}{\mbox{10~ns}} \right)^{0.4}\, \mbox{m\,s${}^{-1}$mJ${}^{-\alpha}$}.
\end{equation}

Several dedicated RALEF simulations have confirmed that the above assumptions and relationships are obeyed with a good accuracy provided that $R_0$ and $t_p$ do not deviate too far from the central values in Eq.~(\ref{app:K_U=}).

\bibliography{Power-law_scaling_Kurilovich_arXiv}

\end{document}